\documentclass[conference]{IEEEtran}

\usepackage{cite}
\usepackage{amsmath,amssymb,amsfonts}
\usepackage{graphicx}
\usepackage{textcomp}
\usepackage{xcolor}
\usepackage{url}
\usepackage{enumitem}
\usepackage{mathtools}
\usepackage{enumitem}
\usepackage{makecell}
\usepackage{multirow, tabularx}
\usepackage[ruled,linesnumbered]{algorithm2e}
\usepackage{support-caption}
\usepackage{subcaption}
\usepackage[justification=centering]{caption}
\usepackage{flushend}

\usepackage{comment}
\usepackage{ifthen}
\newboolean{showcomments}
\setboolean{showcomments}{False}
\ifthenelse{\boolean{showcomments}}
{ \newcommand{\mynote}[3]{
    \fbox{\bfseries\sffamily\scriptsize#1}
    {\small$\blacktriangleright$\textsf{\emph{\color{#3}{#2}}}$\blacktriangleleft$}}}
{ \newcommand{\mynote}[3]{}}
\newcommand{\shrink}[1]{}
\definecolor{dgreen}{RGB}{46, 124, 49}
\definecolor{purple}{rgb}{0.7,0,0.9}

\newcommand{\mh}[1]{\mynote{Mohamed}{#1}{cyan}}

\newcommand{\eg}{\textit{e.g., }}
\newcommand{\ie}{\textit{i.e., }}

\SetCommentSty{algfont}

\begin{document}
\title{\fontsize{23.5pt}{23.5pt}\selectfont ReLeaSER: A Reinforcement Learning Strategy for Optimizing Utilization Of Ephemeral Cloud Resources}

\author{\IEEEauthorblockN{Mohamed Handaoui\IEEEauthorrefmark{1}\IEEEauthorrefmark{3}, Jean-Emile Dartois\IEEEauthorrefmark{1}\IEEEauthorrefmark{2}, Jalil Boukhobza\IEEEauthorrefmark{1}\IEEEauthorrefmark{3}, Olivier Barais\IEEEauthorrefmark{1}\IEEEauthorrefmark{2}, Laurent d'Orazio\IEEEauthorrefmark{1}\IEEEauthorrefmark{2}}
\IEEEauthorblockA{\IEEEauthorrefmark{1}b$<>$com Institute of Research and Technology, 
\IEEEauthorrefmark{2}Univ. Rennes, Inria, CNRS, IRISA.\\ \IEEEauthorrefmark{3}Univ Brest, Lab-STICC, CNRS, UMR 6285, F-29200 Brest. France\\
Email: $\{$mohamed.handaoui, jean-emile.dartois$\}$@b-com.com,\\ boukhobza@univ-brest.fr,
$\{$olivier.barais, laurent.dorazio$\}$@irisa.fr
}
}

\maketitle

\begin{abstract}

Cloud data center capacities are over-provisioned to handle demand peaks and hardware failures which leads to low resources' utilization. One way to improve resource utilization and thus reduce the total cost of ownership is to offer unused resources (referred to as ephemeral resources) at a lower price. However, reselling resources needs to meet the expectations of its customers in terms of Quality of Service. The goal is so to maximize the amount of reclaimed resources while avoiding SLA penalties. To achieve that, cloud providers have to estimate their future utilization to provide availability guarantees. The prediction should consider a safety margin for resources to react to unpredictable workloads. The challenge is to find the safety margin that provides the best trade-off between the amount of resources to reclaim and the risk of SLA violations. Most state-of-the-art solutions consider a fixed safety margin for all types of metrics (\eg CPU, RAM). However, a unique fixed margin does not consider various workloads variations over time which may lead to SLA violations or/and poor utilization.
In order to tackle these challenges, we propose ReLeaSER, a Reinforcement Learning strategy for optimizing the ephemeral resources' utilization in the cloud. ReLeaSER dynamically tunes the safety margin at the host-level for each resource metric. The strategy learns from past prediction errors (that caused SLA violations). Our solution reduces significantly the SLA violation penalties on average by 2.7$\times$ and up to 3.4$\times$. It also improves considerably the CPs' potential savings by 27.6\% on average and up to 43.6\%.

\end{abstract}


\section{Introduction}
\label{section:introduction}
Cloud Providers (CPs) aim to offer resources such as virtual machines or containers with the best Quality of Service (QoS) possible. To do so, data centers are dimensioned according to peak resource usage with the downside of having a low average resource utilization. The low resource utilization increases the Total Cost of Ownership (TCO) which made reclaiming unused resources an urging research topic~\cite{zhang2016history, dartois2018using, javadi2019scavenger}. Resource reclamation is generally made possible thanks to prediction techniques~\cite{zhang2016history, dartois2018using, yang2019elax}. They are usually used to forecast future resource utilization according to customers' behavior in order to infer the unused part (\ie ephemeral resources) and sell it at a lower price.

Customers' workloads (\ie application) running on cloud resources are known to experience sudden variations~\cite{dartois2018using}. They occur due to several factors such as user's request rate and workload types. It causes the resources utilization to increase or decrease in a manner that current predictions cannot always account for. This means that some workload variations are unpredictable or the predictor has failed to discover the hidden patterns~\cite{cao2014cpu, fox2009above}. These sudden variations may cause substantial overestimation or underestimation of resource usage. Overestimation may reduce resource utilization, but underestimation may imply an oversell of resources and thus cause SLA violations and potential cost deficit, which is critical.

In case future resource utilization is unpredictable, a preventive mechanism should be used, such as a safety margin~\cite{yang2017pado, dartois2019cuckoo}. A \textit{safety margin} is a proportion of free resources that are left unused to absorb sudden variations of customers' workloads or predictions' errors in order to guarantee the SLA. The safety margin may be applied at different granularity: a datacenter, a host, or a resource. Choosing the right safety margin value and its granularity is crucial for reducing SLA violations and increasing CP's savings. 

The safety margin may be a static value, that is a fixed proportion of resources applied all the time, for all hosts and resource metrics. This strategy was used in Cuckoo~\cite{dartois2019cuckoo} and Salamander~\cite{handaoui2020salamander} where fixed proportions were empirically tested to select the best one. Although this strategy does decrease potential SLA violations, a substantial amount of resources remained unused due to resource usage overestimations~\cite{handaoui2020salamander}. Moreover, the prediction accuracy of the CPU proved to be lower than that of the RAM~\cite{dartois2018using}, which means that the safety margin should be customized for each resource metric. 
In Scavenger~\cite{javadi2019scavenger}, the authors propose a solution that uses both the mean and standard deviation of the past usage for each resource metric with a fixed sliding window size. Even if this method gives a specific margin for each resource, it requires an additional mechanism to account for a sudden increase in resource utilization.

In this paper, we argue that a dynamic safety margin needs to be employed instead of a static one in order to reduce SLA violation and potentially increase cloud providers' savings. A dynamic solution must consider the following three intrinsic properties of the Cloud environment considered: (1) \textit{volatility} of the resources caused by unpredictable workload changes, (2) \textit{heterogeneity} of the hosts in terms of available resources, and (3) \textit{complexity} of the Cloud dynamics ~\cite{cao2014cpu, fox2009above} that makes it hard to draw an exact model of the variables in play.

Our solution is based on Reinforcement Learning (RL) in order to adjust the size of the safety margin according to the observed prediction errors and violations of customers' SLA. The choice of an RL technique answers the aforementioned properties as follows:
\begin{enumerate}
    \item \textit{Volatility}: the volatility of the reclaimed resources is constantly changing and uncertain. RL is known to be able to reason under such uncertainty~\cite{kaelbling1996reinforcement} and is able to adapt and self-configure with the volatility of resources.
    
    \item \textit{Heterogeneity}: taking heterogeneity of Cloud hosts into account is mandatory since it impacts the performance of workloads. Indeed, RL can be used in order to make decisions for each host separately when properly trained on sufficient data.
    
    \item \textit{Complexity}: Cloud environment cannot be represented with an exact model due to its dynamic and stochastic nature. Thus in many cases, we tend to assume that some variables are known which may impact the performance. However, RL does not require an exact model of the environment in order to learn~\cite{kaelbling1996reinforcement}.
\end{enumerate}

Our strategy, named ReLeaSER, consists of a predictive and reactive approach that dynamically adjusts the safety margin at the host level for each resource metric such as CPU and RAM. In this solution, we suppose that future resource predictions for reclaiming the unused part already exist. The RL solution observes the prediction errors that occurred when using the reclaimed resources and generates the appropriate safety margin. Using the safety margin, we compute both the penalties and the savings for selling the allocated resources.

ReLeaSER was compared to four different strategies for adjusting the safety margin. The comparison was done using real production traces from three datacenters for a 6-months time period. The results show that our solution decreases SLA violations by $2.7\times$ on average as compared to state-of-the-art strategies and increases the CPs savings by $27\%$ on average and up to $43\%$. 


The remainder of this paper is organized as follows. Section~\ref{section:motivation} provides the motivation for using a dynamic safety margin. Section~\ref{section:contribution} details our contribution. Then, Section~\ref{section:experimental_validation} describes the experimental evaluation and the results obtained. In Section~\ref{section:related_work}, we discuss the related work. Finally, Section~\ref{section:conclusion} concludes the paper.

\section{Motivation}
\label{section:motivation}
To motivate our study about the use of a dynamic safety margin, we analyzed two in-production traces for CPU and RAM over a 6-months period. Our study relies on previous work~\cite{dartois2018using} that predicts future resource usage. These predictions are used to reclaim unused resources in order to be allocated to customers. The prediction shows good general accuracy and was used in other studies for scheduling big data applications~\cite{dartois2019cuckoo, handaoui2020salamander}.
\noindent We followed 2 steps in this section:

\begin{enumerate}
\item Using the predictions alongside the in-production traces, we analyzed the prediction errors on different granularities: datacenter, host, resource metric. The intuition behind this analysis is to assess the levels at which the safety margin should be tuned. 

\item We evaluate the CPs' savings when selling the reclaimed unused resources. We also evaluate the impact of prediction errors on SLA violations. Here, we allocate all the resources available according to the predictions of future utilization. The resources are allocated using a single configuration of a container (\ie 2 vCPU, 8 GB RAM) with a price that reflects the real market of volatile resources such as Spot Instance~\cite{amazon_spot_instance}. The prediction errors are computed as follows: $error = prediction - usage$. The total savings are computed with: $overall~savings = savings - SLA~penalties$
\end{enumerate}

To summarize, we focus on three points: 1) the distribution of prediction errors across hosts and resource metrics (\ie CPU, RAM), 2) the CP's savings of the reclaimed resources, 3) the cost of violating SLA guarantees. The setup, datasets used, and cost models are detailed in Section~\ref{section:experimental_validation}. 

\subsection{Prediction errors}
\begin{figure}
    \centering
    \begin{subfigure}[b]{0.24\textwidth}
       \includegraphics[width=1\textwidth]{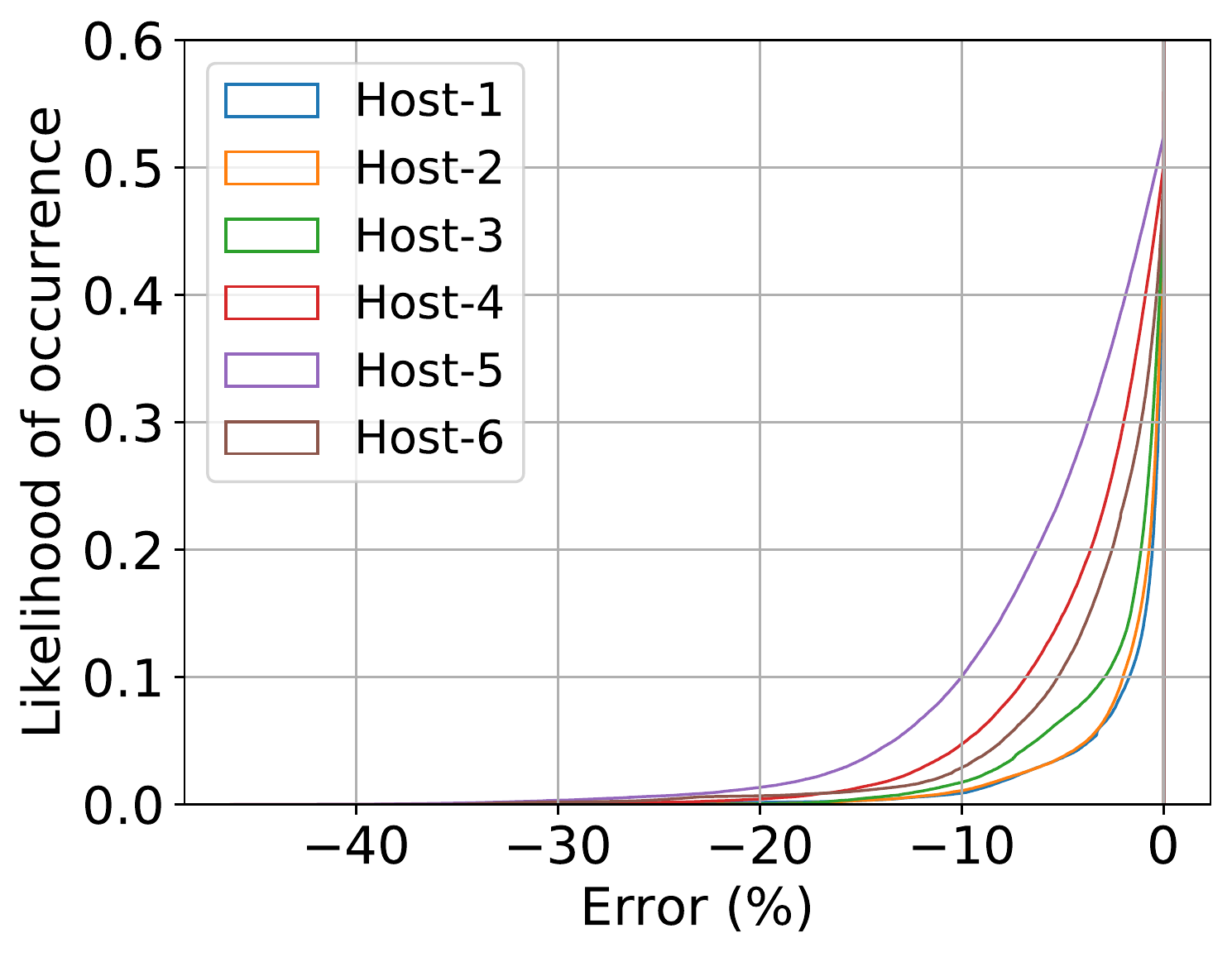}
      \caption{University: CPU errors}
      \label{fig:mot_cpu_cdf_univeristy}
    \end{subfigure}
    \begin{subfigure}[b]{0.24\textwidth}
       \includegraphics[width=1\textwidth]{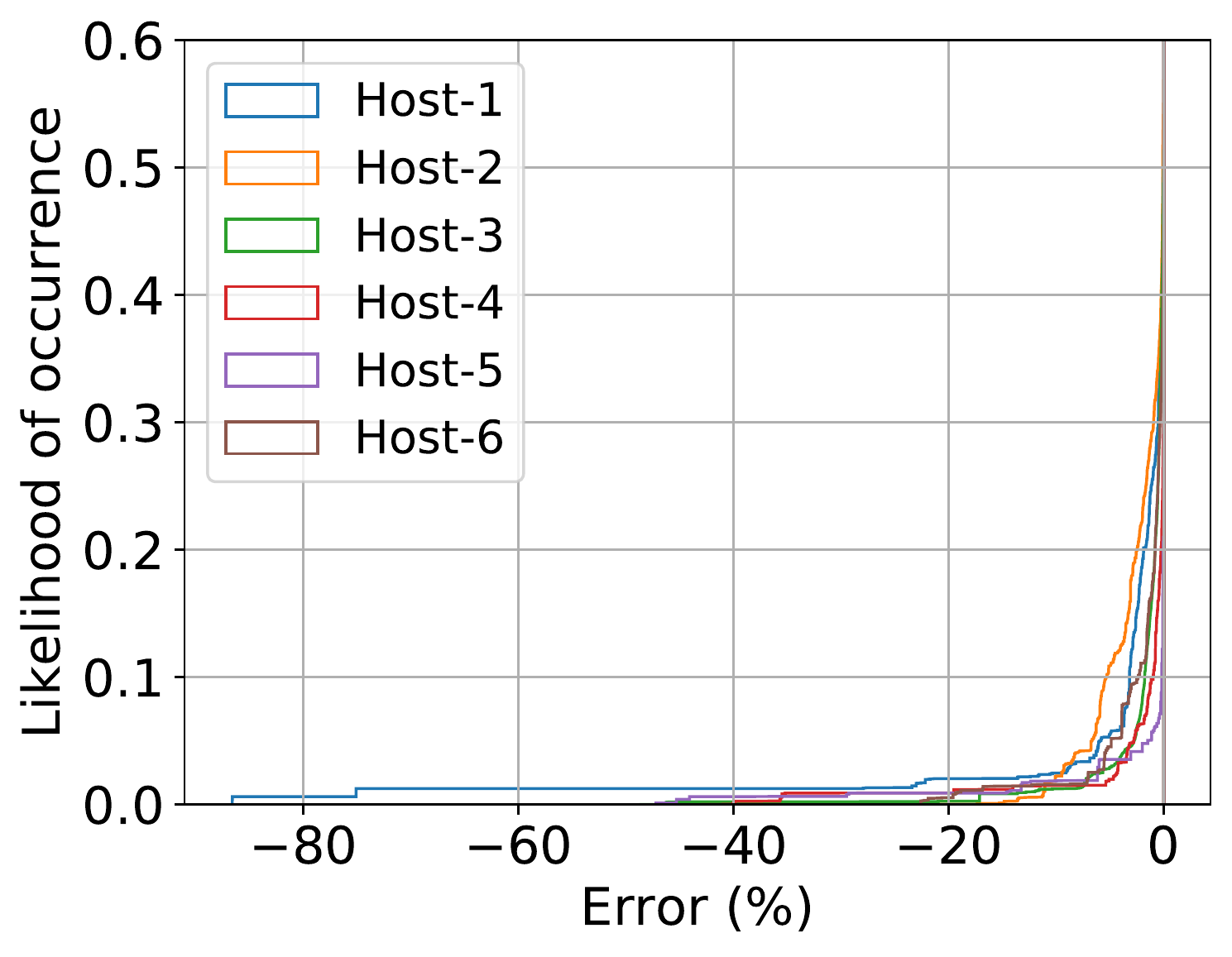}
      \caption{University: RAM errors}
      \label{fig:mot_ram_cdf_univeristy}
    \end{subfigure}

    \begin{subfigure}[b]{0.24\textwidth}
       \includegraphics[width=1\textwidth]{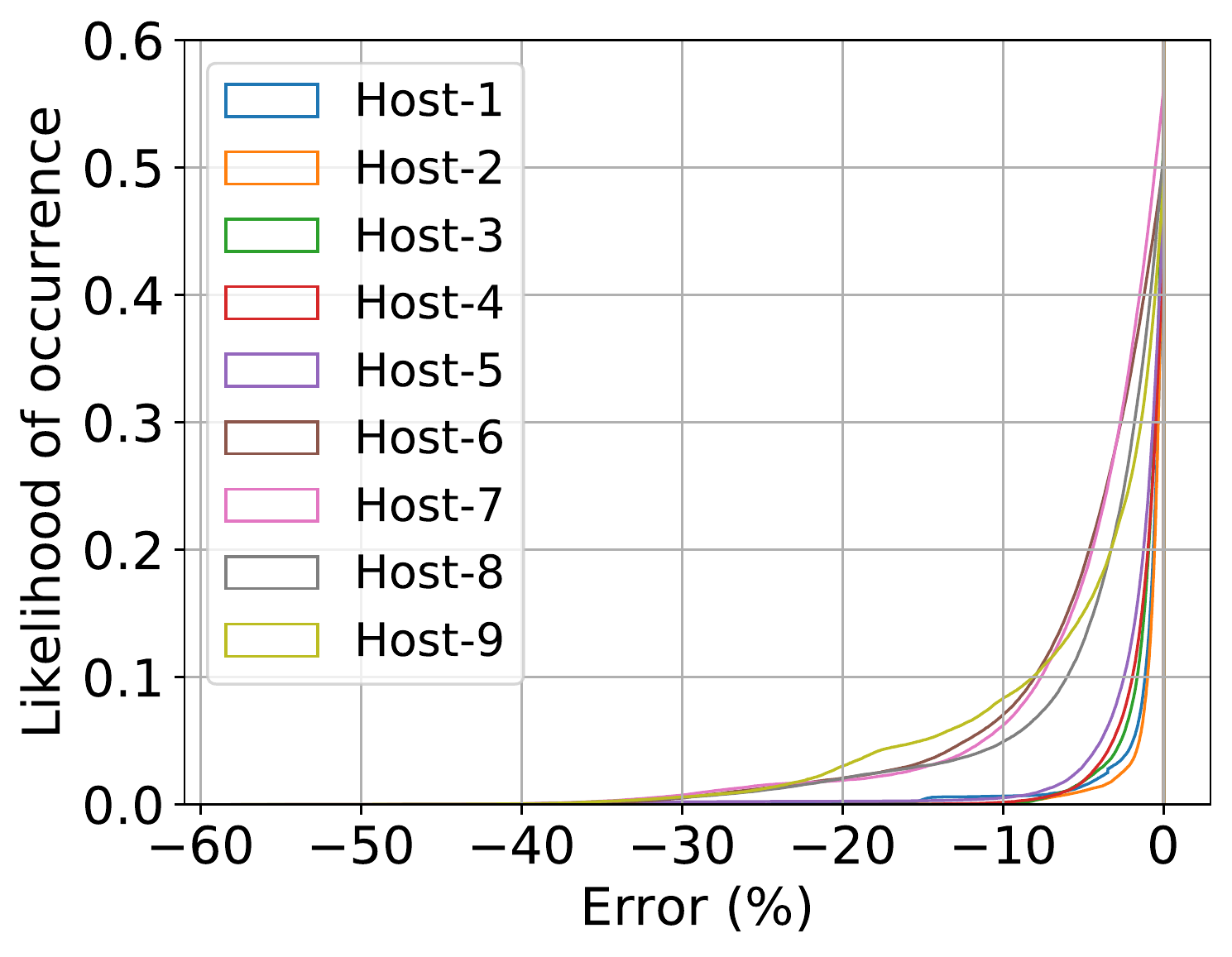}
      \caption{PC-1: CPU errors}
      \label{fig:mot_cpu_cdf_pc1}
    \end{subfigure}
    \begin{subfigure}[b]{0.24\textwidth}
       \includegraphics[width=1\textwidth]{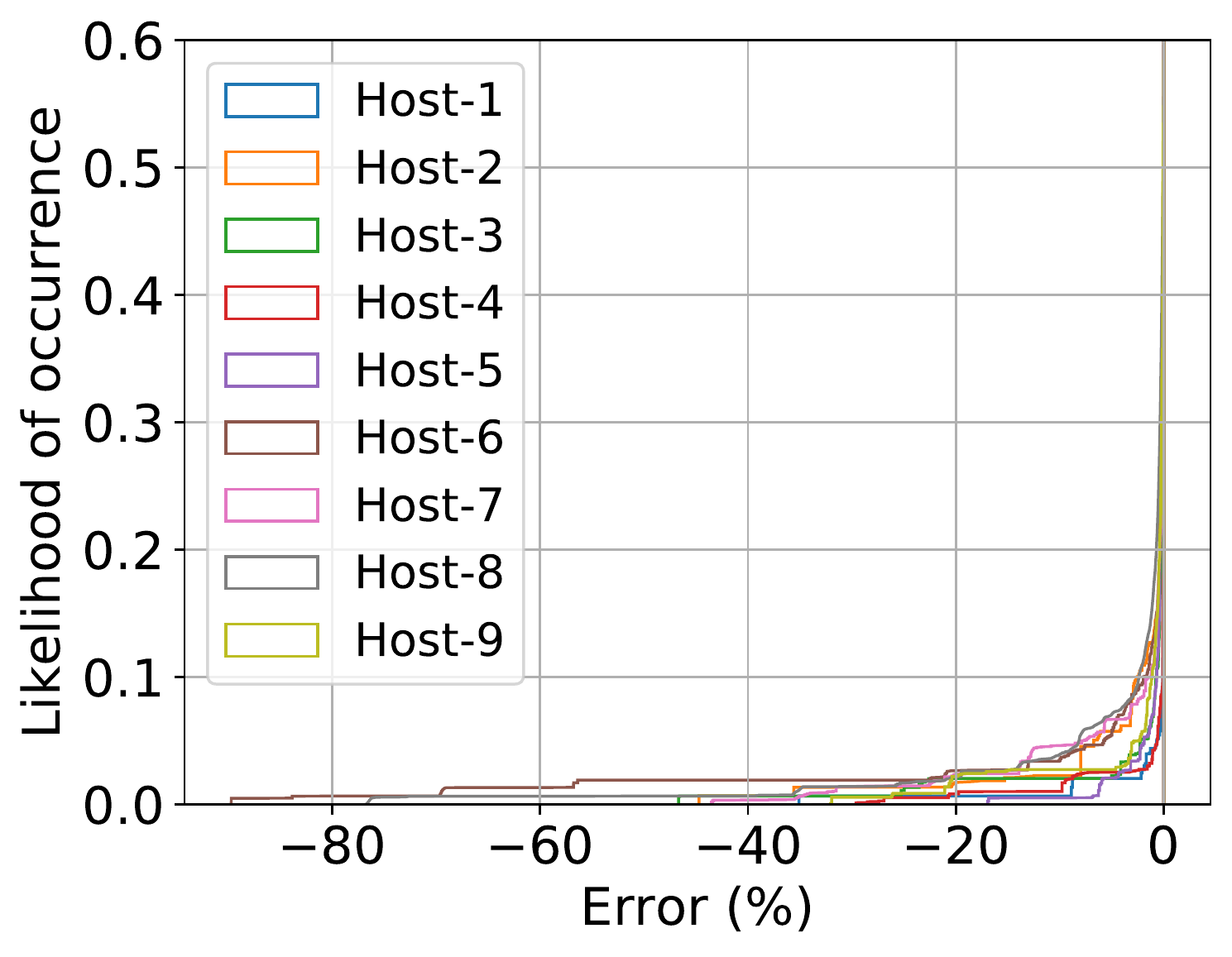}
      \caption{PC-1: RAM errors}
      \label{fig:mot_ram_cdf_pc1}
    \end{subfigure}
    \caption{CDF of resource prediction errors}
    \label{fig:mot_cdf_errors} 
\end{figure}

In this section, we analyze the prediction errors across hosts for both the CPU and RAM. Fig.~\ref{fig:mot_cdf_errors} represents the Cumulative Distribution Function (CDF) of the errors for University and Private Company 1 (PC-1) datasets for a 6-months period. The CDF allows us to estimate the likelihood (\ie percentage) of occurrence of a given prediction error. We only show the underestimation part of the prediction errors as it is the factor causing SLA violation.

When observing the CDF of the RAM in Fig.~\ref{fig:mot_ram_cdf_pc1} and Fig.~\ref{fig:mot_ram_cdf_univeristy}, we notice that the likelihood of prediction errors' occurrence is up to 0.3 (\ie 30\%) for University and 0.2 (\ie 20\%) for PC-1 during the analysis period. We observed that some hosts' prediction errors are as high as 80\% (\eg Host-1 in University, Host-2 in PC-1) but with a lower likelihood of occurrence of around 0.01 (\ie 1\%).
However, when observing the CDF of the CPU in Fig.~\ref{fig:mot_cpu_cdf_pc1} and Fig.~\ref{fig:mot_cpu_cdf_univeristy}, we notice higher errors compared to the RAM. Indeed, for both datasets analyzed, the likelihood of occurrence is around 0.5 (\ie 50\%). This can be explained by the fact that the CPU is less stable thus less predictable~\cite{dartois2018using}.

Next, we compare the different hosts within the same datacenter in each graph of Fig.~\ref{fig:mot_cdf_errors}. We notice that the hosts have different likelihood of occurrence of prediction errors for both CPU and RAM. This is more noticeable for CPU in Fig.~\ref{fig:mot_cpu_cdf_pc1} which is mainly due to the nature of the running workloads on each host. In other words, the workloads in some hosts are more predictable than others.

We noticed that both the observations drawn from the comparison of the CPU and RAM apply to the analyzed traces of the two considered datacenters. The prediction error of resource utilization does change from one datacenter to another. This mainly depends on the type of workloads running in each datacenter and its customers.

From these results, we can conclude that: 1) a safety margin should be set at the host level as some hosts are less predictable than others. 2) the safety margin should be different for the CPU and RAM as the predictability is also not the same. 3) The results were confirmed on three datacenters that we analyzed.

\subsection{Cost related to the reclaimed resources}
\begin{figure}
    \centering
    \begin{subfigure}[b]{0.24\textwidth}
       \includegraphics[width=1\textwidth]{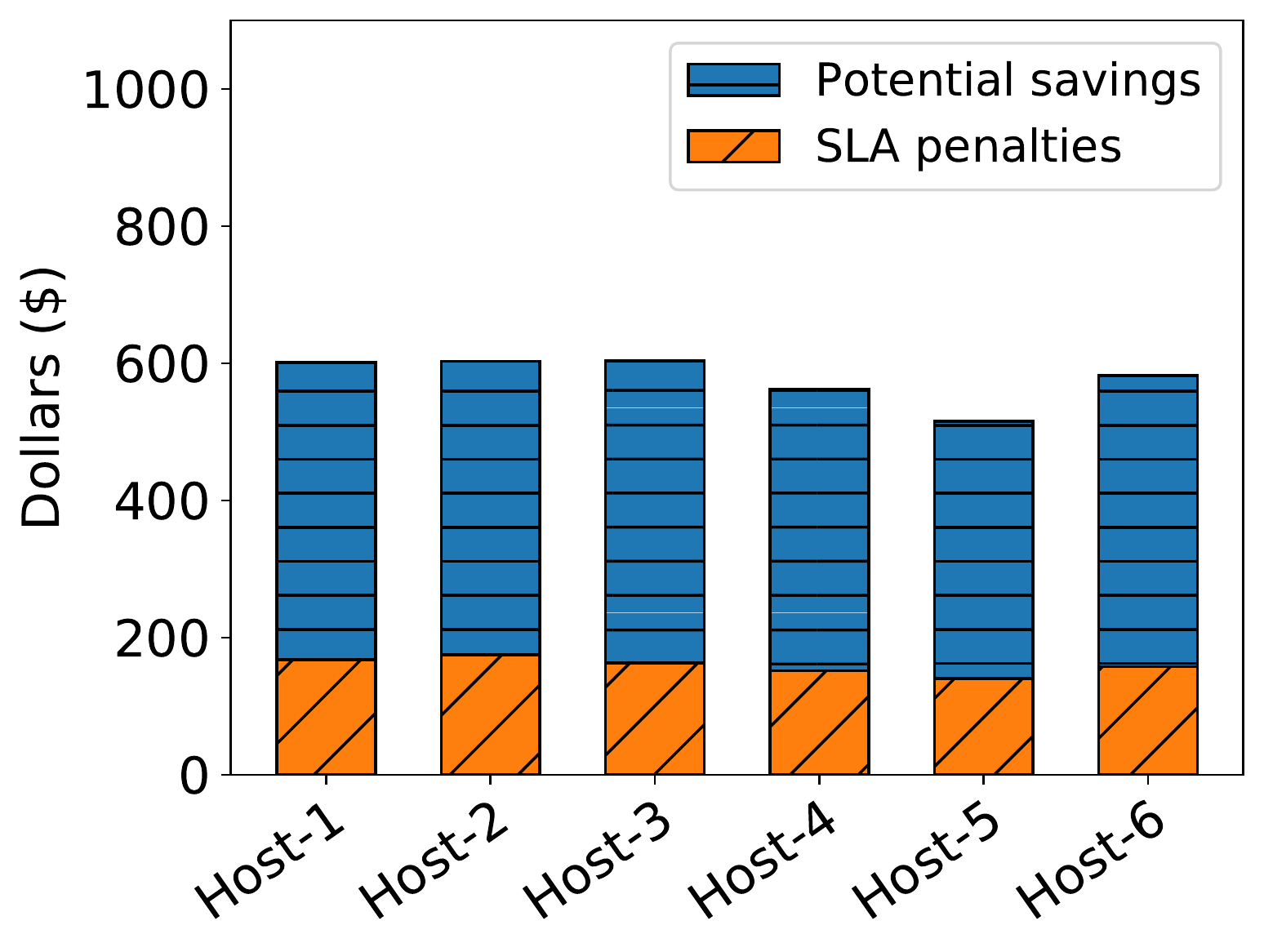}
      \caption{University}
      \label{fig:mot_savings_penalties_university}
    \end{subfigure}
    \begin{subfigure}[b]{0.24\textwidth}
       \includegraphics[width=1\textwidth]{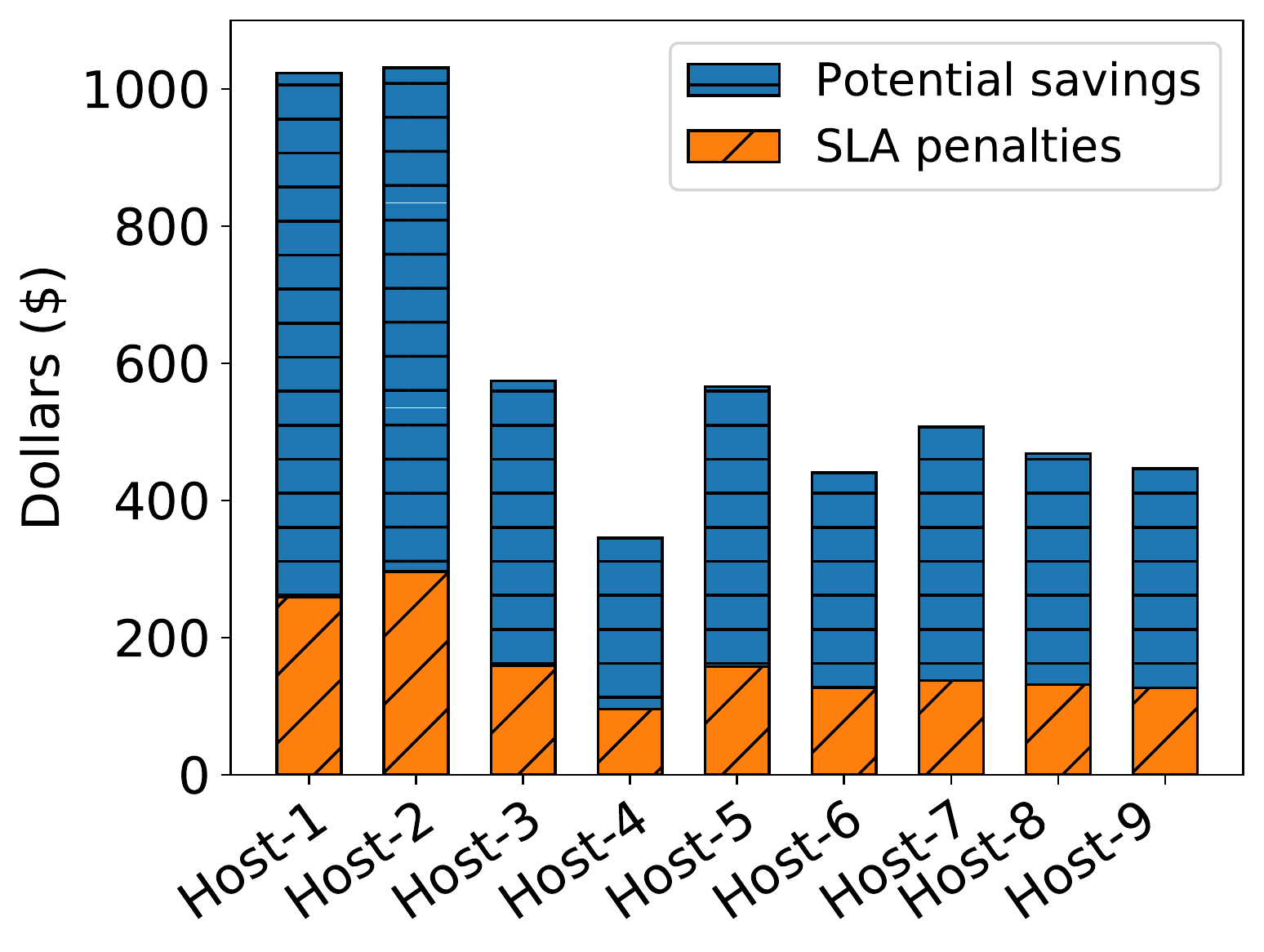}
      \caption{Private Company 1}
      \label{fig:mot_savings_penalties_pc1}
    \end{subfigure}
    \caption{Potential savings and the SLA violation penalties}
    \label{fig:mot_savings_penalties} 
\end{figure}

In the following, we study the economical cost of using the reclaimed resources for both the savings and the SLA violation penalties. The cost model of SLA is detailed in Section \ref{section:experimental_validation}. This model is commonly used by cloud providers such as Google~\cite{google_sla} and Amazon~\cite{amazon_sla}. Fig.~\ref{fig:mot_savings_penalties} represents the potential savings and the SLA violation penalties (in dollars) for each host and during a 6-months period. 

In Fig.~\ref{fig:mot_savings_penalties_pc1}, we observe that the penalties in each host are different mainly due to the heterogeneity of their resources. This confirms the previous statement about setting an appropriate safety margin per host. We also observe that the savings have a similar trend with the penalties. The penalties cost is high when compared to the potential savings of the resources. The total savings in the presented case-study are not huge since the datacenters have a small number of hosts. We can reasonably assume that the savings increase linearly with the number of hosts and become considerable for large datacenters. 

We conclude that resource utilization is not optimized due to the prediction errors that cause SLA violation penalties. Reducing the SLA violations can increase the potential savings. It also increases the reliability of applications. Indeed, SLA violations can have a major impact on the performance and reliability of the running applications. Big data applications, in particular, have to be restarted in case of SLA violation~\cite{yan2016tr, yang2017pado, dartois2019cuckoo}. This highly contributes to wasting resources and may increase the penalties. Thus, providing a dynamic safety margin that is specifically tuned for each host and resource metric may solve the problem.

\section{ReLeaSER: a \textbf{Re}inforcement \textbf{Lea}rning \textbf{S}trategy for optimizing \textbf{E}phemeral Cloud \textbf{R}esources}
\label{section:contribution}
Our goal is to build a solution that optimizes the utilization of unused Cloud ephemeral resources while reducing the risk of violating SLA. This is done by dynamically adjusting the safety margin applied to the resources. The safety margin is a proportion of free resources that are left unused to absorb sudden variations of customers' workloads or predictions' errors\mh{phrase ajouté}. This problem is considered as a control problem since the safety margin is adjusted according to previous errors and potentially other external factors. Note that we aim to dynamically adjust the safety margin to a near-perfect value which should reduce the SLA violations and increase the CPs' savings\mh{phrase ajouté}.

In the following, we present the considered architecture and its modules. Then, we formulate the problem of adjusting the safety margin and the solving algorithm.

\subsection{Architecture overview}
\begin{figure}
    \centering
    \includegraphics[width=0.4\textwidth]{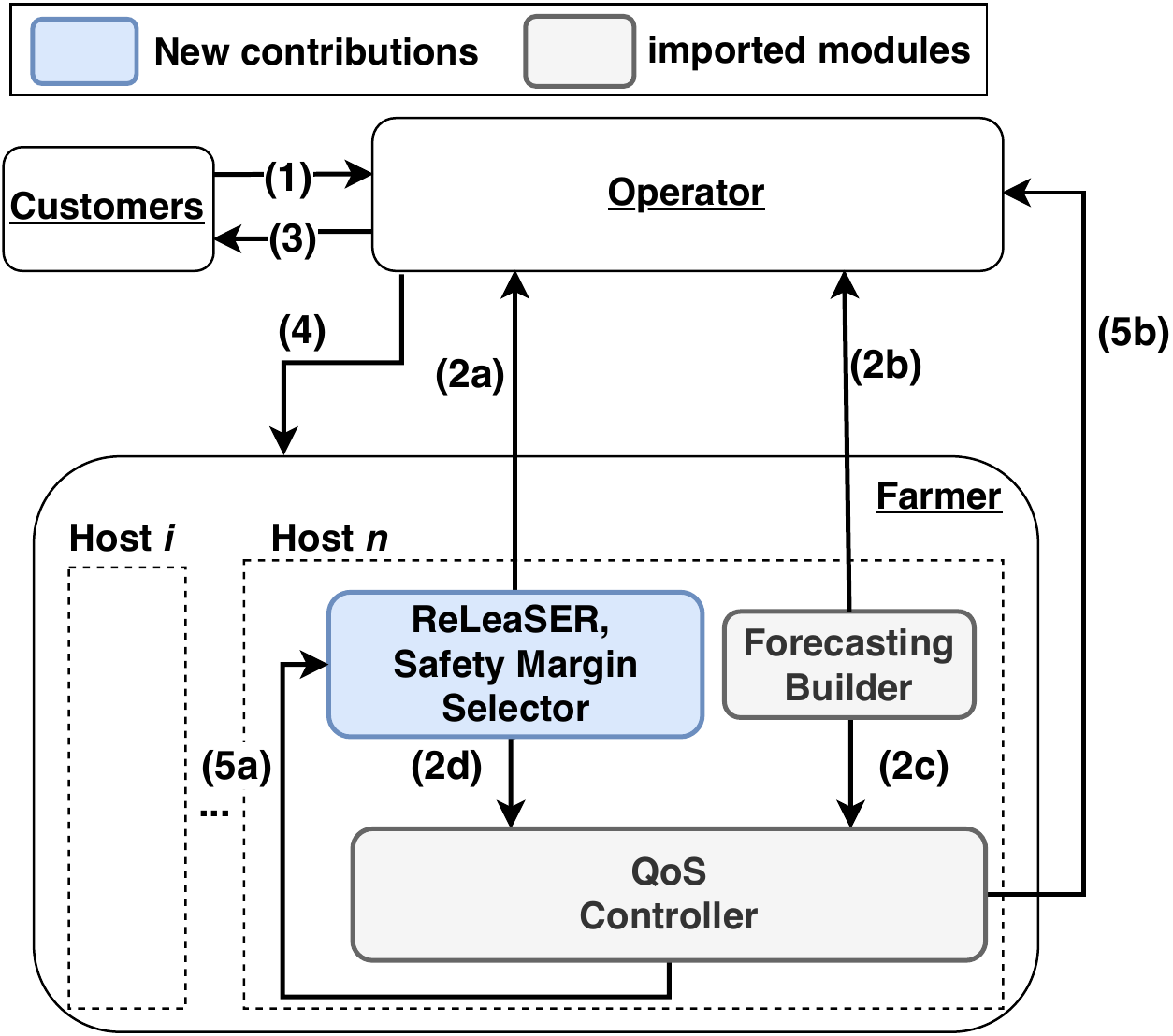}
    \caption{Overview architecture that deploys ReLeaSER: the Margin Selector module}
    \label{fig:architecture} 
\end{figure}

Fig.~\ref{fig:architecture} depicts an overview of the architecture that deploys our Safety Margin Selector module, named ReLeaSER, for adjusting the safety margin. There are three main actors:
\begin{itemize}
    \item \textbf{Farmers}: datacenter owners, they seek to reduce their TCO by offering unused resources to customers. 
    \item \textbf{Customers}: we focus here on customers that request unused volatile resources on the Cloud at a lower cost (we do not consider reserved resources).
    \item \textbf{Operator}: acts as the interface between farmers and customers, they aim at minimizing farmers' TCO by offering unused resources to customers with SLA requirements.
\end{itemize}

\noindent The solution is built upon three main modules (see Fig.~\ref{fig:architecture}):
\begin{itemize}
    \item \textbf{Forecasting Builder}: this module was introduced in~\cite{dartois2018using} and does not constitute a contribution of this paper. The module provides predictions of resource utilization for each host resource metric. This module is not detailed in this paper.
    \item \textbf{QoS Controller}: this module was introduced in previous work~\cite{dartois2019cuckoo, handaoui2020salamander}. 
    We use this module to monitor the utilization of ephemeral resources. It computes the prediction errors considering the safety margins to detect any SLA violations.
    \item \textbf{ReLeaSER, the Safety Margin Selector}: this module represents the core contribution of this paper. It houses the Reinforcement learning algorithm responsible for adjusting the safety margins. It does so by continuously observing the resource prediction errors of the Forecasting Builder (that caused SLA violations) using the QoS controller to act accordingly.
\end{itemize}

In Fig.~\ref{fig:architecture}, the process begins with the customers submitting (1) their request for resources (\ie containers). The operator receives the requests in addition to two other inputs: the predictions of resource utilization (2b) from the Forecasting Builder and the safety margins (2a) from the Safety Margin Selector. The operator decides about which resources can be allocated and sends (3) a response to the customers. In this paper, we do not use a specific container placement algorithm. Since we are evaluating the impact of the safety margin, we suppose that we fully allocate the predicted available resources. That being said, a scheduling component can be easily added when needed as in~\cite{dartois2019cuckoo, handaoui2020salamander}. After that, if the customers' requests can be satisfied, the operator proceeds to allocate (4) the required resources. Meanwhile, the QoS Controller monitors the resource utilization of the host. It checks for underestimated prediction compared to the resource utilization to detect potential SLA violations. To do so, it receives the resource predictions and the safety margins (2c,2d) at the same time as the operator. If any SLA violation is detected, the prediction errors are sent to the Safety Margin Selector (5a) to adjust the future values of the safety margin. The errors are also sent to the operator (5b) in order to act according to the SLA violations.

\subsection{ReLeaSER: the Safety Margin Selector}
In what follows, we describe the Safety Margin Selector module. First, we give some background on Reinforcement learning. Then, we present our problem formulation. Next, we detail the formulation of the reward function (\ie objective function). Finally, we describe the solving algorithm. 

\subsubsection{Reinforcement Learning}
RL is an area of machine learning~\cite{kaelbling1996reinforcement} that can be used to solve problems that require a series of decisions. The algorithm learns what action to do so as to maximize a numerical reward signal. The algorithm is not told which actions to take (from the predefined set of actions), but instead must discover which ones yield the highest reward by trying them. RL is based on Markov Decision Process (MDP)~\cite{kaelbling1996reinforcement}. MDP is a discrete-time stochastic control process~\cite{bellman1957markovian}. It offers a mathematical framework for modeling problems where the results are sometimes random and sometimes under the control of a decision-maker.

The Main concepts of an MDP are the following:
\textit{i) Agent}: the decision-maker that sets the size of the safety margin, 
\textit{ii) Environment}: a Cloud host which is the interface that an agent interacts with,
\textit{iii) State}: it describes the environment (\ie host) properties at a given time which can be observed by the agent.
\textit{iv) Action}: is what an agent can do in each state (\ie change the size of the safety margin),
\textit{v) Reward}: is a feedback signal from the environment to the agent (\ie SLA violation cost, allocated resources cost).

The objective of solving an MDP is to find the optimal policy $\pi$ (a function that specifies the action to take for each state) that maximizes the sum of expected future rewards.
\vspace{.2cm}

\subsubsection{Problem formulation}
we formulate the problem of adjusting the safety margin by using the MDP framework with the tuple $\{S, A, R,\mathbb{P}\}$. It formulates the state of the environment, the action of our agent, the reward function and states transitions:

\begin{itemize}
    \setlength\itemsep{0.5em}
    \item  $S = \{ errors  \} $, $S$ represents the current state of the environment (\ie host). The state indicates the previous prediction $errors$ during a predefined time window. The $errors=[e(t-w, m), ... ,e(t, m)]$ is a sliding window vector of size $w$ where each value is computed using: $e(t, m) = u(t, m) - p(t, m)$. 
    Where $e(t, m)$, $u(t, m)$ and $p(t, m)$ represent the prediction error, the host real utilization and the prediction respectively for resource metric $m$ at time $t$.
    The size of the error window $w$ was set to one hour in order to have a reactive strategy that adapts quickly to workload changes.
 
    \item $A =\{ sm \in \mathbb{R} \;|\; 0 \leq sm < 100 \}$ is the action set which consists of the possible percentage values for the safety margin. The safety margin is generated for each time step $t$. The time step is set to 3 minutes similarly to the prediction sampling. This allows for the algorithm to adjust quickly since the prediction error changes at each step.

    \item $ \mathbb{P}: S \times A \times S \to [0,1]$ is the probability that the environment transitions from state $s$ to a new state $s'$ when action $a$ is performed (\eg an increase in resource utilization when placing a container). MDP modeling requires this variable, but in our case, we used RL algorithms which implicitly consider these transitions~\cite{arulkumaran2017deep}. Indeed, due to the complexity of the cloud environment, it is hard, if not impossible, to model precisely its state transitions.
    
    \item $R : S \to \mathbb{R}$ is the cost function expressing the expected reward when the system is in state $s$. The reward function does not depend on the action $a$ (\ie safety margin) as SLA penalties at $t$ are not necessarily a result of the immediate previous action ($t-1$). They are also due to mispredictions which may occur at any moment. Thus, the reward signal is not immediate but delayed according to the applied safety margin at $t$. The reward function is detailed below.
\end{itemize}

\subsubsection{Reward function}
in what follows we detail the reward function. The idea here is to reward the agent when allocating resources but penalize it in case of SLA violation. Thus the reward function can be formulated according to the total savings of resources while considering SLA penalties: \mh{j'ai modifié la notation: $c_{total\_savings}$ $\rightarrow$ $c_{savings}$ et $c_{savings}$ $\rightarrow$ $c_{potential\_saving}$} 
\begin{IEEEeqnarray}{lCr}
c_{savings}(h,d) = c_{potential\_saving}(h, d) - c_{penalty}(h, d)~~~~~
\end{IEEEeqnarray}

\noindent With $c_{savings}(h,d)$ representing the savings for a given host $h$ and day $d$. $c_{potential\_saving}(h, d)$ is the potential savings when no SLA violation occurs from allocating the available resources in host $h$ and day $d$. $c_{penalty}(h, d)$ represents the penalties due to SLA violations in host $h$ during the day $d$. 

The potential savings $c_{potential\_saving}(h, d)$ of a Cloud provider for the allocated resources is formulated as follows:
\begin{IEEEeqnarray}{lCr}
c_{potential\_saving}(h, d) = \sum_{t \in 24h} nb_{container}(h, d, t) * ppm~~~
\end{IEEEeqnarray}

With $nb_{container}(h, t)$ being the number of containers in a host $h$ during $t$. $ppm$ represents the price per minute of hosting a container. The price depends on the size of the allocated container and its price per hour $pph(container\_size)$.

The penalty of SLA violations $c_{penalty}(h, d)$ is computed using a discount percentage which is deduced according to the duration of violation in a 24-hour window (\eg see Table~\ref{tab:sla_discounts}):
\begin{multline}
    c_{penalty}(h, d) = c_{potential\_saving}(h, d) \; * \\
                        discount(T_{violation}(h, d))
\end{multline}

Where $discount(T_{violation}(h, d))$ is the discount percentage according to the measured duration of violating SLA (\eg see Table~\ref{tab:sla_discounts}). The time duration $T_{violation}(h, d)$ of violating SLA is incremented every time step for which a violation is observed:
\begin{IEEEeqnarray}{lCr}
\text{\textbf{if}  } p(h, d, t) < u(h, d, t) \text{  \textbf{then}  } T_{violation}(h, d) ~+= ts
\end{IEEEeqnarray}
With $p(h, d, t)$ being the prediction of the resource usage $u(h, d, t)$ for host $h$ during day $d$ at time $t$. $ts$ represents the time step in minutes.

\subsubsection{The solving algorithm} one of the main criteria for choosing an RL algorithm is the type of the action space~\cite{kaelbling1996reinforcement} (\ie discrete or continuous), in our case, the safety margin. On the one hand, a discrete action algorithm outputs an action from a finite set of possible actions. This means that for our problem, we would need to discretize the safety margin space. If we choose a 5\% step, we would need to train the algorithm with 20 actions (from 0 to 100\%) which can be time-consuming~\cite{dulac2015deep}. On the other hand, a continuous action algorithm outputs an action with real values. This means that one action as the output would suffice as it can be any real value in the interval $[0,100]\%$ for the safety margin percentage.

\begin{figure}
    \centering
    \includegraphics[width=0.35\textwidth]{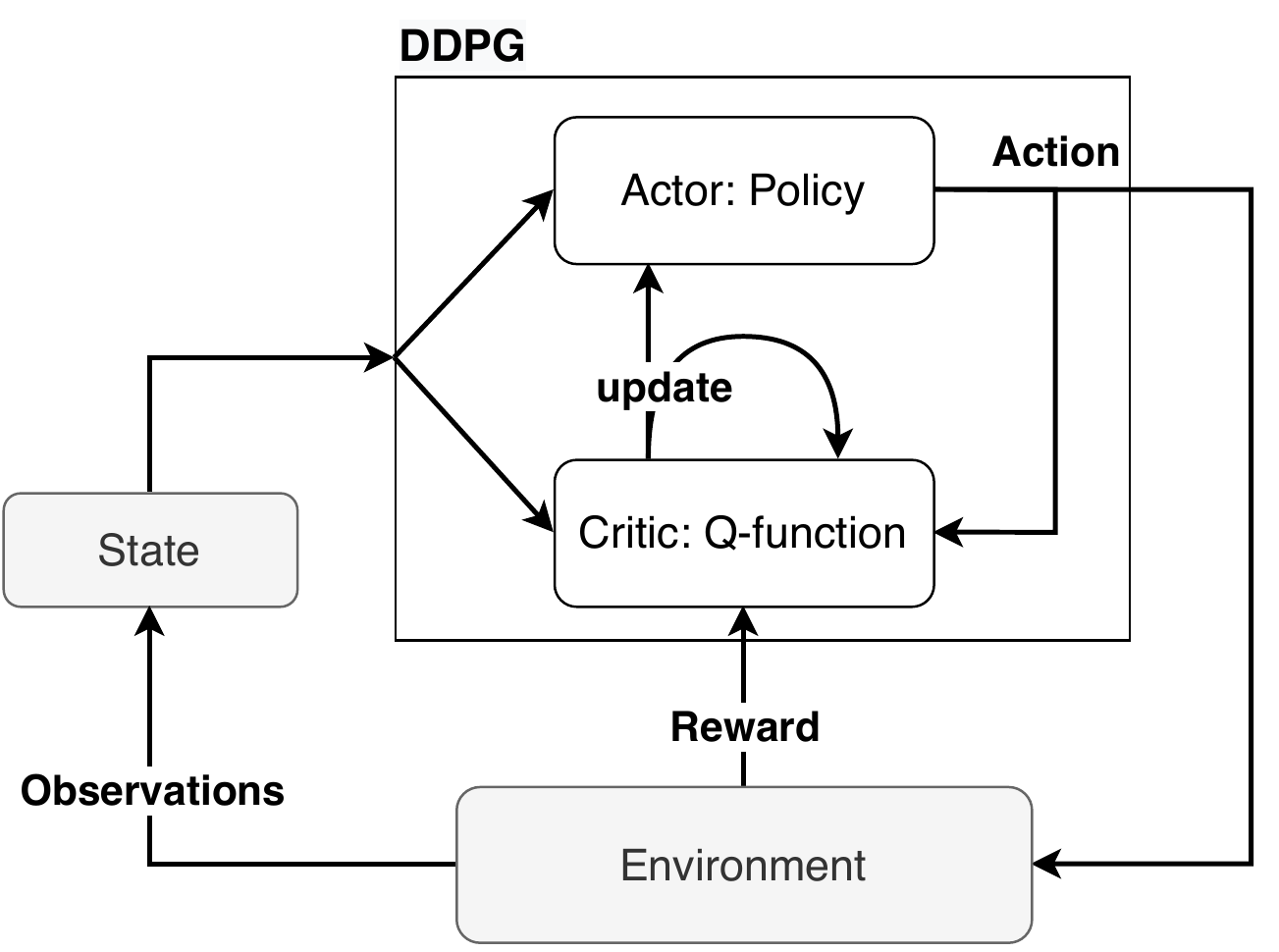}
    \caption{Overview diagram of the DDPG algorithm} \label{fig:ddpg_overview} 
\end{figure}

We chose one of the state-of-the-art algorithms~\cite{arulkumaran2017deep} called Deep Deterministic Policy Gradient (DDPG)~\cite{lillicrap2015continuous} which is efficient for the continuous action space problem~\cite{henderson2018deep}. DDPG is a Reinforcement learning algorithm that concurrently learns a Q-function (\ie the value that represents the quality of state-action pairs) and a policy (\ie the mapping of states to actions).
DDPG uses two neural networks called Actor and Critic which are represented in Fig.~\ref{fig:ddpg_overview}. The actor is used to learn the policy (\ie choosing a value of the safety margin), whereas the critic computes the Q-function value of the actor's action which is used in updating the networks.

Using the DDPG algorithm, we can integrate the formulated problem such as the observation of prediction errors, selection of safety margins, and the reward function. 
Algorithm.\ref{algo:ddpg} represents the pseudo-code used for configuring the safety margin during the learning and testing phases. 

First, we initialize two variables (lines 1-2) used by DDPG: i) the discount factor $\gamma$ is used to balance the importance of immediate and future rewards. We set the value to $\gamma=0.99$ which prioritizes future rewards. ii) The learning rate $\alpha$ is used in machine learning algorithms. It should balance the convergence accuracy of the algorithm and the learning speed. Lower values slow down the learning but improve the convergence of the agent. We empirically selected $\alpha=0.001$ by evaluating different values that reduce the learning speed for better convergence. 

We then initialize a replay buffer (line 3) that stores previous experience for faster convergence during the training. Then, we create the DDPG model (line 4) which contains both the actor and critics networks.
The algorithm loops over the traces by days then by a predefined time step. In each step, we get the resource prediction (line 7) and its usage (line 8). We compute the prediction errors using the previously computed safety margin (line 9). The errors are then used to compute the reward function (line 10).

The DDPG agent uses the observed prediction errors to select a safety margin (line 11). Initially, the algorithm does not have any experience. Thus, a random process must be used to make random actions for exploration purposes. For efficient learning, we have to balance between exploration (\ie searching for new knowledge) and exploitation (\ie improving upon the current knowledge). In the function executed to select a safety margin (line 11), we used Ornstein–Uhlenbeck process~\cite{doob1942brownian}. It is an algorithm used for the exploration/exploitation problem in the case of continuous action space. Finally, when training the algorithm, we store the previous experience in the replay buffer (line 13). From the replay buffer, we randomly select a batch of previous experiences in order to update the agent's model.

\SetNlSty{textbf}{}{:}
\begin{algorithm}
    \SetAlgoLined
    \SetNoFillComment

    $\alpha=0.001$; \tcp{learning rate}
    $\gamma=0.99$; \tcp{discount factor}
    experience = initializeReplayBuffer()\;
    agent = DdpgModel($\alpha, \gamma$)\;
    
    \For(\tcp*[h]{loop over days}){d = 1, D}{
        \For{t = 0, 24h; time step}{
            reward = 0\;
            \For{h \textbf{in} hosts}{
                predictions = getResourcePrediction(h, d, t)\;
                usages = getResourceUsage(h, d, t)\;
                errors = prediction + sm - usage\;
                reward += computeRewardValue(errors)\;
            }
            sm = agent.selectSafetyMargin(errors)\;
            \If{training}{
                experience.store(errors, sm, reward)\;
                \If{updateRequired()}{
                    batch = experience.randomSamples()\;
                    agent.update(batch)\;
                }
            }
        }
    }
  \caption{Pseudo-code of configuring the safety margin using DDPG}
  \label{algo:ddpg}
\end{algorithm}

\section{Experimental validation}
\label{section:experimental_validation}
In this section, we detail the experimental setup and results used to validate the efficiency of our contribution and try to answer the following questions: 

\begin{itemize}
    \item[\textbf{Q1:}] What is the overall performance of ReLeaSER compared to other strategies in terms of savings and SLA penalties?
    
    \item[\textbf{Q2:}] What are the potential gains of ReLeaSER on larger production datacenters?
    
    \item[\textbf{Q3:}] How was the safety margin adapted for each datacenter/host/resource metric?
\end{itemize}

\vspace{.2cm}

\textbf{Experiment metrics and strategies:} 
Comparing the proposed solution to other strategies is realized using the same metrics presented in the Motivation Section~\ref{section:motivation} namely: 1) the cost of SLA violation, 2) the total savings related to the reclaimed resources. Both of these metrics are used to assess the quality of the selected safety margin. Finding a trade-off between SLA violation penalties and the total savings determines the performance of the strategy.
ReLeaSER is compared to the following strategies:
\begin{itemize}
    \item \textbf{Random}: this strategy sets the safety margin randomly. It was evaluated to observe whether our strategy effectively learns rather than choosing random actions.
    \item \textbf{Fixed}: this strategy empirically selects the best safety margin for all host and resources. It was used in~\cite{dartois2019cuckoo, handaoui2020salamander}, the best safety margin value for the tested solutions and datacenters was 5\%.
    \item \textbf{Simple feedback}: this strategy simply adds to the safety margin of 5\% from the fixed strategy, the prediction error from the previous time step.
    \item \textbf{Scavenger}: this strategy uses both the mean of the resource usage during a time window and the standard deviation to build an interval of future utilization. The value of the standard deviation can be used as a safety margin. Scavenger~\cite{javadi2019scavenger} was used to reduce interference between applications.
\end{itemize}


\subsection{Implementation}
ReLeaSER was implemented using Keras-rl~\cite{plappert2016kerasrl} v.$0.4.2$ that implements state-of-the-art deep reinforcement learning algorithms in Python. It is based on Keras~\cite{chollet2015keras} v.$2.3.1$, a framework used to develop deep machine learning models. Keras is built on top of Google's open-source framework TensorFlow~\cite{abadi2016tensorflow}. We used TensorFlow GPU v.$1.14.0$. The configuration of additional parameters is required~\cite{plappert2016kerasrl}:
\begin{itemize}
    \item Replay memory (number of steps): $limit = 100000$.
    \item Ornstein–Uhlenbeck process: $size=1$, $theta=0.15$, $mu=0$, $sigma=0.3$.
    \item Number of warm-up steps (actor/critic): $1000$.
    \item Batch size: $128$.
    \item Error metric: Mean Absolute Error (MAE): \vspace{-.2cm}
    $$MAE=\frac{1}{n}\sum\limits_{j=1}^{n}(y_j-\hat{y}_j)$$
    With $y_j$ is the target value and $\hat{y}_j$ is the observed value.
    \item Training/Testing ratio: $training = 0.8$, $testing = 0.2$. 
    \item Target model update: after 10 windows of 24 hours.
\end{itemize}

\mh{ajouté}A neural network~\cite{goodfellow2016deep} is comprised of neurones in layers namely \textit{Input layer}, \textit{Output layer} and all intermediate layers are called \textit{Hidden layers}. Layers are interconnected with a specific type of connection such as \textit{dense} where all neurons of two layers are fully connected. Finally, each layer has an activation function that controls the output. The neural network architecture of the agent is similar to the DDPG Pendulum example of Keras-rl~\cite{ddpg_pendulum_example}:
\begin{itemize}
    \item Actor's architecture:
    \begin{itemize}
        \item \textit{Input layer}: dense layer, 10 neurons (state input size), ReLu activation,
        \item \textit{Hidden layers}: two dense layers, 16 neurons, ReLu activation,
        \item \textit{Output layer}: dense layer, one neuron (action), Linear activation.
    \end{itemize}
    \item Critic's architecture:
    \begin{itemize}
        \item \textit{Input layer}: dense layer, 11 neurons (state input size + Actor's action), ReLu activation,
        \item \textit{Hidden layers}: two dense layers, 32 neurons, ReLu activation,
        \item \textit{Output layer}: dense layer, one neuron (Q-value of the action), Linear activation.
    \end{itemize}
\end{itemize}

To train the algorithm, we split the dataset (called PC-2) into 80\% for the training phase and 20\% for the testing. The split is done on the 6-months period of the dataset comprising 27 hosts. We used PC-2 dataset because it has the highest number of hosts. After the training, random actions are not performed but only the learned strategy to assess exactly the performance of the algorithm. We also evaluate two additional datasets of PC-1 and University.

\begin{table}
    \renewcommand{\arraystretch}{1.3}
    \centering
    \caption{Total capacities and average resource \\utilization of datacenter~\cite{dartois2018using}\mh{ajouté average CPU and RAM utilization}}
    
    \begin{tabular}{ |c|c|c|c|c|c| }
        \hline
        \thead{Datacenter} & \thead{Number\\of hosts} & \thead{CPU \\ (cores)} & \thead{Average \\ CPU usage} & \thead{RAM \\(TB)} & \thead{Average \\ RAM usage} \\ 
        \hline
        PC-1 & 9 & 120 & 14.6\% & 1.2 & 55.7\%\\
        \hline
        PC- 2 & 27 & 230 & 10.3\% & 3.8 & 43\% \\
        \hline
        University & 6 & 72 & 9.8\% & 1.5 & 60.4\%\\
        \hline
    \end{tabular}
    \label{table:data-centers-capacities}
\end{table}

\subsection{Experimental setup}
A summary of the datasets is presented in Table~\ref{table:data-centers-capacities}, it shows the overall capacity and average utilization of all datacenters which are heterogeneous. PC-1 (\ie Private Company 1) has 6 different configurations among its 9 hosts. PC-2 has 13 different configurations among its 27 hosts. University has 6 different configurations. More details can be found in~\cite{dartois2018using}.
In order to compute the potential savings, we used the following models in~\cite{dartois2018using}:
\begin{itemize}
    \item Leasing model: a unique model is used for simplicity which is a container with 2 vCPUs, 8 GB RAM.
    \item Pricing model: a fixed price for the leasing model of one container. It is based on a pay-as-you-go model. The price was fixed to 0.0317\$/hour as Amazon Spot Instance~\cite{amazon_spot_instance}. 
    \item Penalty model: a delay-dependent penalty of SLA violations for which the discount is relative to the CP response delay. Table~\ref{tab:sla_discounts} shows the discounts applied according to the accumulated time of SLA Violations during a day.
\end{itemize}

\begin{table}
    \renewcommand{\arraystretch}{1.3}
    \centering
    \caption{Discount percentage in case of\\ violations during a 24-hour day~\cite{dartois2018using}}
    \begin{tabular}{|c|c|}
        \hline
        Violation Duration (Minutes) &  Discount\\
        \hline
        $>$ 15 to $\leq$ 120 & 10\%\\ \hline
        $>$ 120 to $\leq$ 720 & 15\%\\ \hline
        $>$ 720 & 30\%\\
        \hline
    \end{tabular}
    \label{tab:sla_discounts}
\end{table}

\subsection{Experiment results}

\begin{figure*}[htbp]
    \centering
    \begin{subfigure}[b]{0.32\textwidth}
      \centering
      \includegraphics[width=0.9\textwidth]{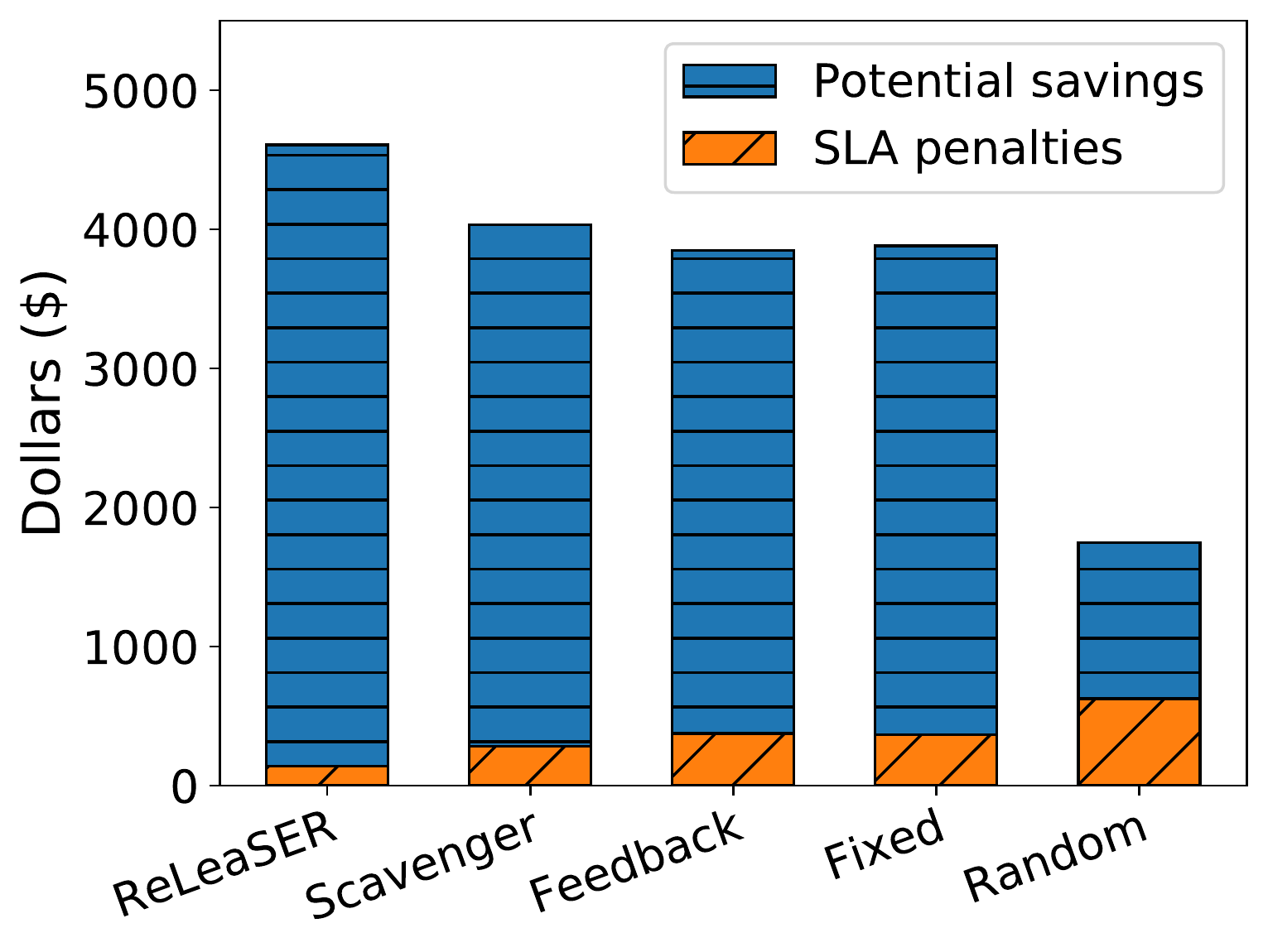}
      \caption{Private Company 1}
      \label{fig:comparison-cost-pc1}
    \end{subfigure}
    \begin{subfigure}[b]{0.32\textwidth}
      \centering
      \includegraphics[width=0.9\textwidth]{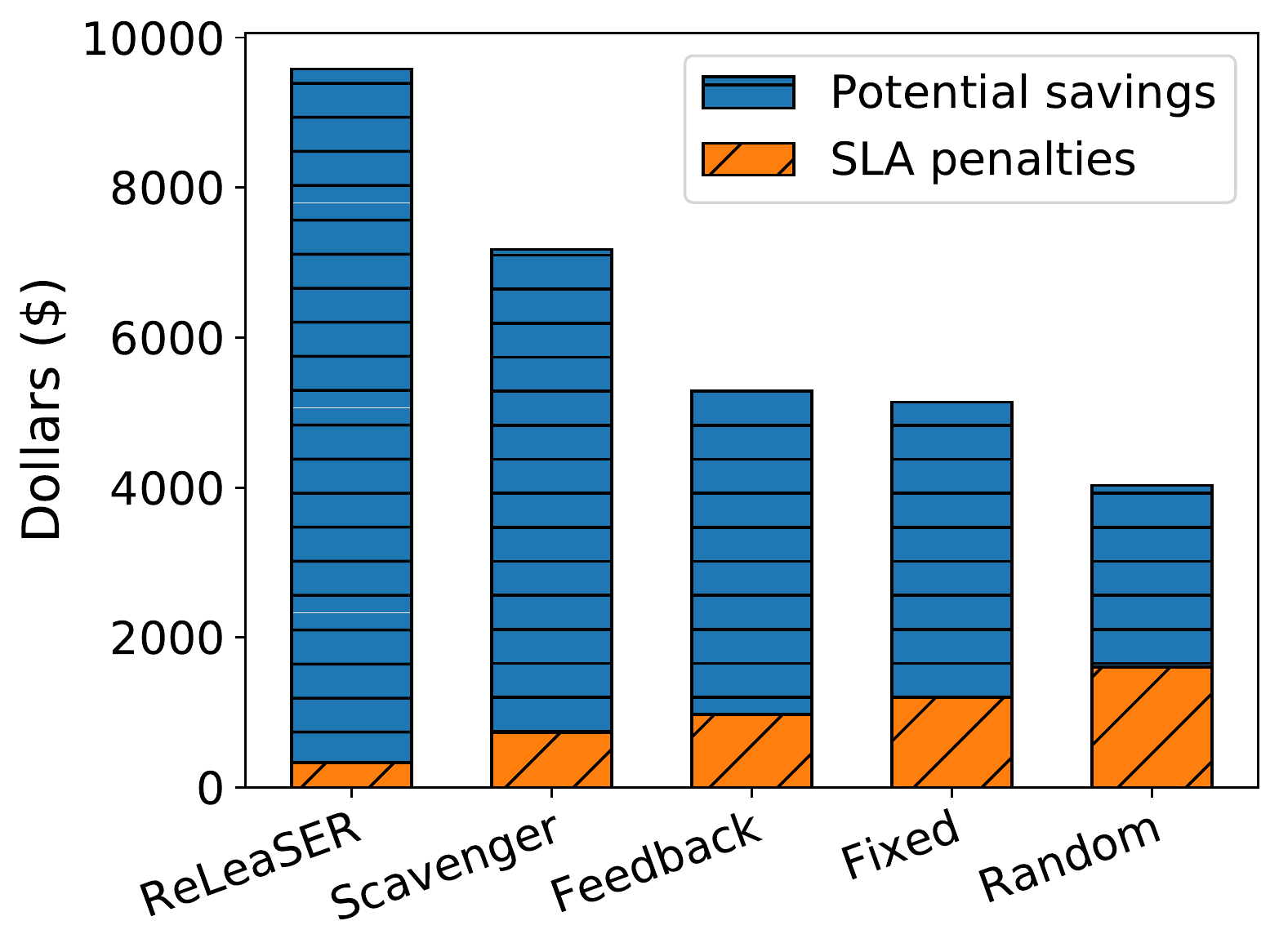}
      \caption{Private Company 2}
      \label{fig:comparison-cost-pc2}
    \end{subfigure}
    \begin{subfigure}[b]{0.32\textwidth}
      \centering
      \includegraphics[width=0.9\textwidth]{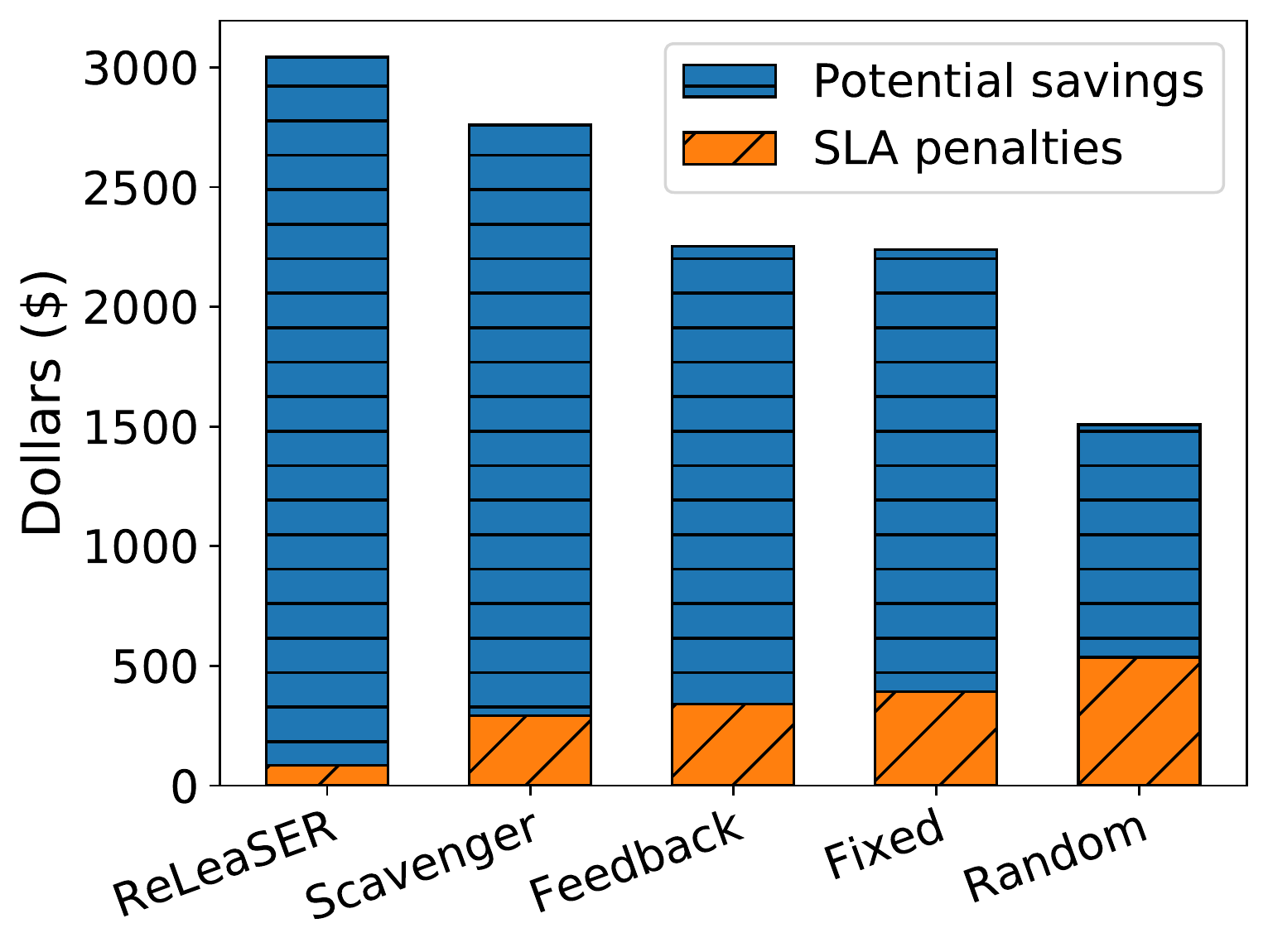}
      \caption{University}
      \label{fig:comparison-cost-university}
    \end{subfigure}
    
    \caption{Comparison of the overall savings and SLA violation penalties of the reclaimed resources}
    \label{fig:comparison-cost}
\end{figure*}

\subsubsection{Q1-Cost of allocating resources} in this experiment, we evaluate the cost of allocating the reclaimed resources. We compare both the overall savings of CPs and the SLA violation penalties.
Fig.~\ref{fig:comparison-cost} are stacked-bar graphs that represent the SLA penalties (orange) and the overall savings (blue) for the different strategies and datacenters. 

\textbf{SLA penalties}: when observing the SLA penalties, we notice that all the strategies are able to reduce the penalties compared to the values seen in Section~\ref{section:motivation}. This is expected as long as the value of the safety margin is greater than zero. Among the strategies, the random one performed the worst. Then the fixed one followed by the simple feedback strategy. The latter performed better because it considers a minimum value for the safety margin. However, the top two performing strategies are ReLeaSER with the least SLA violations penalties then Scavenger. Indeed, when comparing the improvements of ReLeaSER to Scavenger, it reduces penalties on average by $2.7\times$ ($1.8\times$, $2.9\times$, $3.4\times$ for PC-1, PC-2, and University respectively). 

\textbf{Overall savings}: when observing the overall savings, we note that the random strategy also performs the worst since it has the highest violation rate. Both the fixed and simple feedback have comparable savings in spite of the difference in SLA violation. This can be explained by the size of the selected safety margin. Choosing a larger safety margin does reduce SLA penalties but does not necessarily improve savings. A trivial example that showcases this is a safety margin of 100\% that leads to no penalties but also no savings. 

When comparing ReLeaSER to Scavenger, we observe an improvement in the overall savings by $27.5\%$ on average. Our strategy improves savings by $43.6\%$, $19.1 \%$, $19.8 \%$ corresponding to PC-1, PC-2, and University respectively. However, we can notice that although ReLeaSER is up to 3x better than Scavenger when it comes to penalties reduction, the savings were improved by up to $43\%$. This highly depends on the penalty model used, as one model can be more penalizing than another.

\subsubsection{Q2-Extrapolation to larger datacenters} the previous evaluation was done on relatively small datacenters compared to what Amazon and Google offer. Hence, the savings computed on the previous experiment allow only for an objective comparison. In here, we extrapolate the savings for both ReLeaSER and Scavenger on Amazon datacenters configuration.

Each Amazon datacenter has between 50000 and 80000 hosts~\cite{king2016augmented}. We computed the approximate savings for a datacenter with 50000 hosts by using the average savings of each strategy. With an average saving of $\sim\!\!82$\$ per host and per month for ReLeaSER and $\sim\!\!65$\$ for Scavenger, we obtained the following results: When using Scavenger on an Amazon datacenter, the savings of the reclaimed resources is $3250000\$$/month. Whereas, the savings using ReLeaSER is $4100000\$$/month. This means that our solution increases the total potential savings by 21\% or by $850000\$$ per month compared to Scavenger. Even though the extrapolation may seem naive, it gives a rough idea about the savings that can be achieved.

\begin{table}
    \renewcommand{\arraystretch}{1.3}
    \setlength{\tabcolsep}{5pt}
    \centering
    \caption{Safety margins for University hosts}
    \begin{tabular}{ |c|c|c|c|c|c|c| } 
     \hline
      & Host-1 & Host-2 & Host-3 & Host-4 & Host-5 & Host-6 \\
     \hline
     Minimum & 0\% & 0\% & 0\% & 1\% & 2\% & 1\% \\
     \hline
     Median  & 3\% & 2\% & 4\% & 5\% & 6\% &  4\%\\
     \hline
     75th percentile & 4\% & 4\%  & 7\% & 10\% & 18\% & 9\% \\
     \hline
    \end{tabular}
    \label{table:sm_univ_hosts}
\end{table}

\subsubsection{Q3-Analysis of the selected safety margins}
\begin{figure}
    \centering
    \begin{subfigure}[b]{0.24\textwidth}
      \centering
      \includegraphics[width=\textwidth]{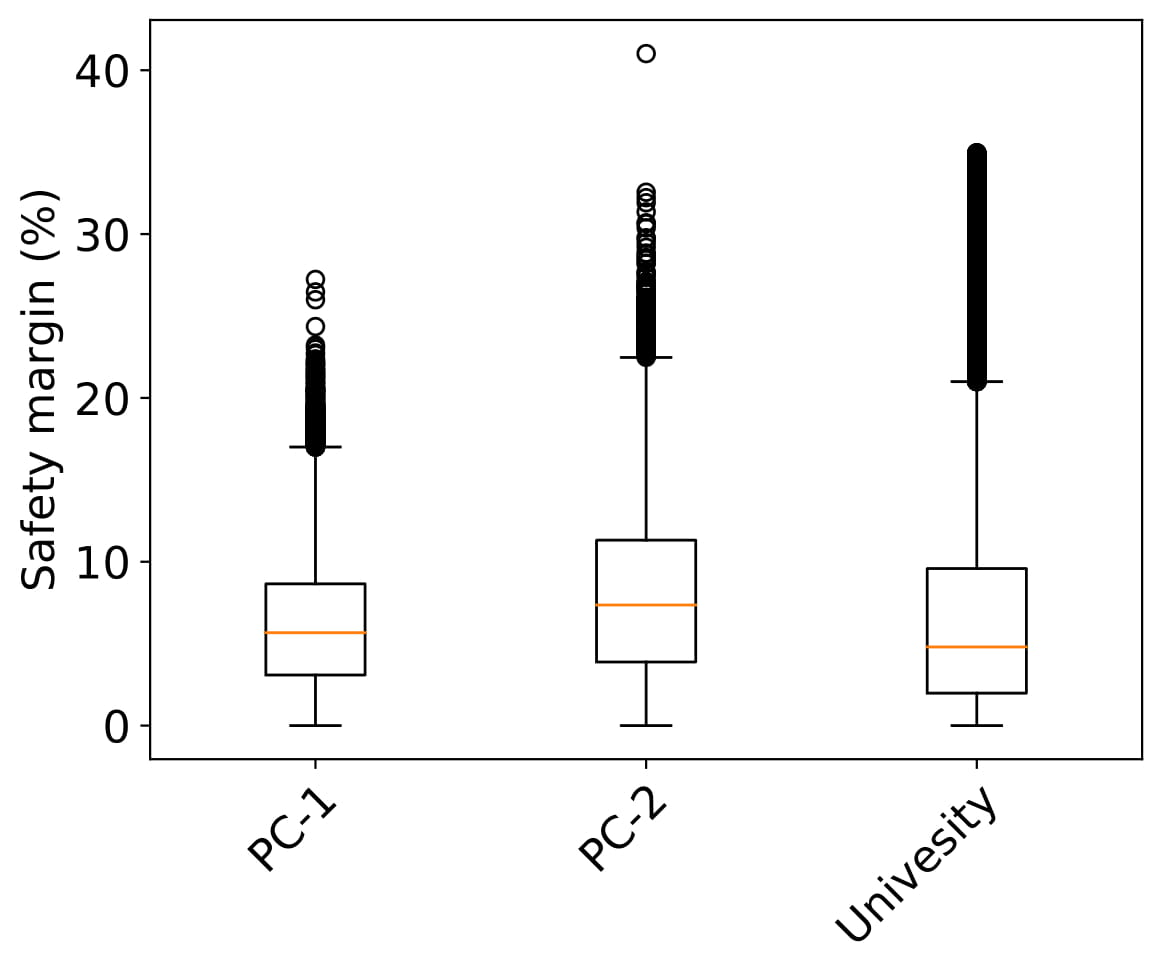}
      \caption{CPU safety margins}
      \label{fig:comparison-sm-cpu}
    \end{subfigure}
    \begin{subfigure}[b]{0.24\textwidth}
      \centering
      \includegraphics[width=\textwidth]{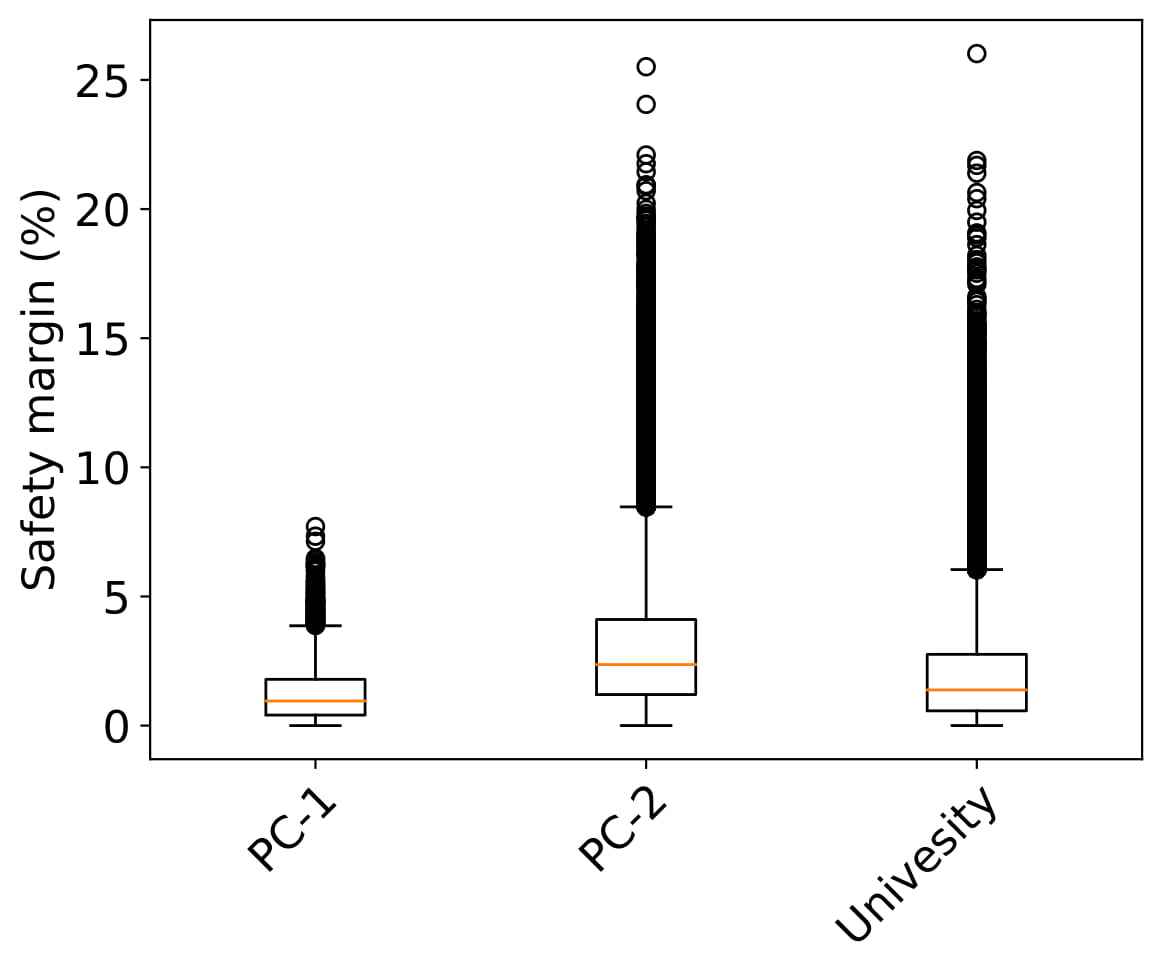}
      \caption{RAM safety margins}
      \label{fig:comparison-sm-ram}
    \end{subfigure}
    \caption{Safety margins selected by ReLeaSER}
    \label{fig:comparison-sm}
\end{figure}

in this section, we analyze the safety margins selected by ReLeaSER. The goal is to understand how it performed and where do the gains come from. Fig.~\ref{fig:comparison-sm} represents boxplot graphs of the selected safety margins for the different datacenters for both the CPU (Fig.~\ref{fig:comparison-sm-cpu}) and the RAM (Fig.~\ref{fig:comparison-sm-ram}). Table~\ref{table:sm_univ_hosts} shows the minimum, median and 75th percentile of the safety margins for University hosts. 

The first observation that can be drawn from Fig.~\ref{fig:comparison-sm} is that the median safety margins in the case of CPU are higher than the RAM's. This confirms that the RAM should be tuned separately as seen in Section~\ref{section:motivation}. Moreover, the median safety margin of the CPU is around 5\% which aligns perfectly with the best safety margin of the fixed strategy. Each datacenter has different values of the safety margin used in order to reduce the penalties. We observe that the safety margins change from one datacenter to another with PC-2 giving the highest value. This difference demonstrates that each datacenter has different behavior in terms of resource utilization. 
Finally, we observe that there are some outliers for all datacenters considered (see Fig.~\ref{fig:mot_cdf_errors}) which are due to high prediction errors. However, their likelihood of occurrence is low.

In Table~\ref{table:sm_univ_hosts}, we observe the different safety margin levels set for each host. The minimum, median, and 75th percentile values vary from one host to another. Host-5 has the highest median and the 75th percentile of the safety margins. While Host-1 and Host-2 have the lowest similar safety margins which may mean that they both have similar predictability. These results confirm that the safety margin should be tuned at a host-level.

\section{Related work}
\label{section:related_work}
The safety margin of resources is used in a variety of applications (also referred to as headroom). In big data applications, such as Hadoop~\cite{shvachko2010hadoop}, a user-configurable safety margin can be used for each host. This safety margin, however, is mainly used for decisions such as re-prioritizing sub-tasks to take advantage of currently allocated containers.
In Pado~\cite{yang2017pado}, Cuckoo~\cite{dartois2019cuckoo} and Salamander~\cite{handaoui2020salamander}, a fixed safety margin is used. As shown in the motivation Section~\ref{section:motivation}, the safety margin should be configured for each host and resource metric since they exhibit different behaviors.
In \cite{zhang2016history}, the authors propose a safety margin tailored to the job execution time. The higher the execution time of the job, the larger is the safety margin. However, this technique is specific to big data jobs. This means that, if multiple jobs are executed, the safety margin is set to the longest job even if most jobs have a short execution time.
Rhythm~\cite{zhao2020rhythm} and CLITE~\cite{patel2020clite} are two frameworks used for optimizing resource utilization by co-locating latency-critical applications. Rhythm uses a load limit which is the upper limit of request load to allow the co-location. The number of co-located applications is controlled by the lower limit of request load called slack. Similarly, CLITE computes a maximum request load by evaluating the latency for each ran application. CLITE also evaluates the maximum load of memcached that guarantees the QoS requirements. However, both Rhytm and CLITE need to build a catalog of applications that can be co-located which is limiting and time-consuming to extend. Our strategy, ReLeaSER, adjusts the safety margin dynamically without specifying the type of the running workloads. Instead, it relies only on the host resource utilization and its prediction to reduce SLA violations and increase savings.

\section{Conclusion}
\label{section:conclusion}
Using reclaimed resources is important for Cloud providers in order to increase their savings. However, allocating reclaimed resources should be done while guaranteeing customers SLA which is challenging. In addition, resource reclamation may rely on prediction mechanisms that are error prone in view of the stochastic nature of Cloud workloads.

On account for these challenges, we propose ReLeaSER, a Reinforcement Learning strategy for optimizing the utilization of ephemeral resource in the cloud. The strategy consists in setting a dynamic safety margin on a host-level basis for each resource metric. The strategy learns from the prediction errors and improves the size of the safety margin accordingly. This is done to reduce the SLA violation penalties and increase the potential savings of Cloud providers.

We evaluated ReLeaSER with four other strategies for adjusting the safety margin. The results show that we reduce considerably the SLA violation penalties on average by 2.7 times. The improvements are also considerable for the CP's savings with an average of 27.5\%. Furthermore, ReLeaSER can save approximately $4100000\$$/month when linearly extrapolated to a single Amazon datacenter.

For future work, we plan to extend our work to additional resource metrics such as network and storage. We also plan to evaluate the strategy with higher volatility of resources. Also, we did not consider the starting time variations of the containers and virtual machines. This may have an impact on the relevance of the chosen strategy, which might reduce the efficiency of ReLeaSER. We plan to upgrade our agent model to consider such starting time fluctuations. \mh{ajouté} Finally, we plan to implement Safe Reinforcement Learning~\cite{garcia2015comprehensive} which is used to avoid random actions. This can be useful for giving the RL agent the chance to improve and adapt its strategy while in production without impacting the performance.

\section*{Acknowledgment}
This work was supported by the Institute of Research and Technology b-com, dedicated to digital technologies, funded by the French government through the ANR Investment referenced ANR-A0-AIRT-07.

\bibliographystyle{IEEEtran}
\bibliography{references}
\end{document}